\documentclass[lettersize,journal]{IEEEtran}
\usepackage[T1]{fontenc}
\usepackage{amsmath,amsfonts}
\usepackage{algorithmic}
\usepackage{algorithm}
\usepackage{array}
\usepackage[caption=false,font=normalsize,labelfont=sf,textfont=sf]{subfig}
\usepackage{textcomp}
\usepackage{stfloats}
\usepackage{url}
\usepackage{verbatim}
\usepackage{graphicx}
\usepackage{cite}
\usepackage{multirow}
\usepackage{multicol}
\usepackage{booktabs}
\usepackage{xcolor}
\usepackage{soul}

\newcommand\thefont{\expandafter\string\the\font}

\begin{document}

\title{Reinforcement Learning for Freeway Lane-Change Regulation via Connected Vehicles}

\allowdisplaybreaks
\author{Ke Sun,~\IEEEmembership{Student Member, IEEE} and Huan Yu, ~\IEEEmembership{Senior Member, IEEE}
\thanks{Ke Sun and Huan Yu are with the Hong Kong University of Science and Technology (Guangzhou), Thrust of Intelligent Transportation, Guangzhou, Guangdong, China. This work was supported by the National Natural Science Foundation of China No.12526214, No.62203131, and Guangzhou Municipal Education Bureau University Project 2024312102. Huan Yu is the corresponding author (Email: {\it huanyu@hkust-hz.edu.cn}).}}

\maketitle
\begingroup
\renewcommand{\thefootnote}{}
\footnotetext{© 2026 IEEE.  Personal use of this material is permitted.  Permission from IEEE must be obtained for all other uses, in any current or future media, including reprinting/republishing this material for advertising or promotional purposes, creating new collective works, for resale or redistribution to servers or lists, or reuse of any copyrighted component of this work in other works.}
\endgroup

\begin{abstract}
Lane change decision-making is a complex task due to intricate vehicle-vehicle and vehicle-infrastructure interactions. Existing algorithms for lane-change control often depend on vehicles with a certain level of autonomy (e.g., autonomous or connected autonomous vehicles), which limits their applicability under low penetration rates of automated vehicles. To address this issue, this study proposes a lane-change regulation framework using multi-agent reinforcement learning (MARL) to improve freeway traffic efficiency via connected vehicles. Specifically, lane-change regulation signals (e.g., `left change allowed'' or `right change disallowed'') are computed at a traffic management center and then broadcast to connected vehicles. Thus, human-driven vehicles remain uncontrolled, while connected vehicles follow the received regulation signals. Compared to vehicle-level maneuver control, this approach reduces communication and positioning requirements, avoids direct trajectory intervention, and can accommodate different levels of connectivity and autonomy. To further bridge the gap between theoretical macroscopic analysis and microscopic simulations, the proposed framework combines a microscopic simulation environment and a macroscopic lane-grid representation. The microscopic simulator is used to generate vehicle-level trajectories and execute the realized traffic dynamics, while the macroscopic framework aggregates microscopic vehicle information into lane-grid states, enabling low-cost control at the grid-agent level. Based on a multi-lane macroscopic traffic model represented by partial differential equations (PDEs), lane changes are interpreted as source-term exchanges between adjacent lanes, which can be regulated by the proposed lane-change regulation signals. Our model is evaluated in varied traffic scenarios and demand conditions using microscopic simulation. Experimental results demonstrate that the method improves overall traffic efficiency with limited additional energy consumption while maintaining driving safety.

\end{abstract}


\section{Introduction}

Freeway traffic control has relied on infrastructure-based measures such as speed limits and ramp traffic lights. Recently, there has been a growing focus on leveraging connected automated vehicles (CAVs) for traffic flow improvement. The cruising behaviors of the CAVs are controlled to enhance traffic stability \cite{cui2017stabilizing,stern2018dissipation,talebpour2016influence,wang2023deep}, safety \cite{zhao2023safety,zhao2024safety,chen2024safety}, energy consumption \cite{rios2018impact,jin2018experimental,vahidi2018energy}, and efficiency \cite{vcivcic2021coordinating,milanes2013cooperative,talebpour2016influence,hyldmar2019fleet}. While significant advancements have been achieved in longitudinal vehicle and platoon control, lateral control—particularly in lane change control—remains a challenging task due to the complexity of optimization in a higher-dimensional solution space.

Many studies focus on ramp-merging areas in freeway traffic control, where mandatory lane changes of ramp-merging vehicles can cause substantial disturbances to mainline traffic. Traditionally, macroscopic approaches such as ramp metering have been employed to control the inflow from on-ramps. Classical methods include the feedback control strategy ALINEA \cite{papageorgiou1997alinea} and its extensions \cite{zhang2001evaluation,smaragdis2004flow,papageorgiou1990modelling}. More recently, Deep Reinforcement Learning (DRL)-based approaches have been explored to achieve superior performance under model uncertainty and incomplete information \cite{belletti2017expert,yu2021reinforcement,fares2014freeway,deng2019advanced}. With advances in connectivity and autonomy technologies, CAVs offer an opportunity to control individual vehicles in ramp-merging areas instead of relying solely on aggregated control. A classical method is to map the merging problem to a platoon control framework using virtual mapping \cite{milanes2010automated,chen2021connected,liao2021cooperative}. Beyond this mapping framework, optimization-based methods have been proposed to search for optimal solutions \cite{cao2015cooperative,zhou2018optimal,chen2020hierarchical}. DRL-based methods have also been introduced \cite{guo2020merging,zhang2023high}. 

Another critical challenge involves managing lane changes on freeways to improve traffic efficiency, which is the primary focus of our work. Studies such as \cite{pan2016modeling,xu2011analysis} demonstrate that lane changes significantly impact traffic safety and efficiency. Further research shows that CAVs can enhance these metrics in traffic congestion, particularly with high CAV penetration rates \cite{wang2019review,monteiro2023safe}. At the macroscopic level, an implicit strategy for managing lane changes is the implementation of restricted-use lanes, such as High Occupancy Vehicle (HOV) lanes \cite{menendez2007effects,naseri2024simulating}. Related multi-class macroscopic traffic models further show the importance of class-aware and lane-level representations. Classical multi-class extensions of LWR-type and kinematic-wave models describe heterogeneous traffic by assigning different states or behavioral parameters to different vehicle classes \cite{wong2002multi,logghe2008multi}. More recent multiclass CTM-based models describe mixed traffic with human-driven vehicles and autonomous or connected automated vehicles, where traffic capacity, wave speed, and lane-level flow dynamics depend on the class composition and penetration rate \cite{levin2016multiclass,pan2021multiclass}. To optimize individual lane change decisions and trajectories, researchers have developed collaborative driving strategies. For example, game-theory-based methods, such as the Nash bargaining approach in \cite{lin2019pay}, aim to optimize traffic flow. Optimization-based methods \cite{tajalli2022distributed,xu2020bi} seek to mitigate congestion near conflict areas by coordinating CAV lane change decisions. Additionally, Multi-Agent Reinforcement Learning (MARL) methods have been proposed to achieve robust generalization performance \cite{bhalla2020deep,chen2023deep}.

However, existing DRL models for lane change control face two significant challenges. First, given the expected low penetration of CAVs in the near future, performance under varying penetration rates must be carefully evaluated. Experimental findings in \cite{talebpour2016influence} reveal that while CAVs can enhance traffic stability, substantial improvements are primarily observed at higher penetration rates. Second, the accurate collection of traffic state information, particularly in large-scale environments, is often prohibitively expensive or even infeasible. Many existing approaches assume robust vehicle-to-vehicle or vehicle-to-infrastructure communication. However, as noted in \cite{hua2023multi}, challenges such as communication latency, limited data bandwidth, and privacy concerns can impede the effectiveness of centralized training schemes.

To address these deficiencies, we propose a traffic flow control strategy within a multi-agent reinforcement learning (MARL) framework. This strategy regulates lane-changing behaviors of connected vehicles (CVs) at the macroscopic level to enhance the overall efficiency of mixed freeway traffic. Inspired by infrastructure-based control methods such as ramp metering and restricted-use lanes for freeway management, our model employs an infrastructure-based regulation framework that operates at the decision level. Regulation signals\textemdash either permitting or prohibiting maneuvers\textemdash are dispatched through a traffic management center to CVs. Consequently, our algorithm follows the paradigm of centralized training with decentralized execution (CTDE). A shared policy is trained in the traffic management center, and each agent selects its own actions independently.

The minimal regulation units in our model are spatially discretized cells of the macroscopic PDE model, referred to as lane grids. Lane-grid-based methods have become increasingly common in recent lane-level variable speed limit (VSL) control.
Lane changes often reduce the effectiveness of VSL strategies, especially near traffic bottlenecks. To address this issue, lane grids are preferred over aggregated road segments because they capture inter-lane speed differences more accurately \cite{carey2015extending}. In \cite{lu2023td3lvsl, lu2024improving}, experimental results show that fine-resolution VSL control outperforms traditional road-segment-based approaches. In \cite{guo2020integrated}, mandatory lane-change control is combined with lane-grid VSL, demonstrating superior efficiency compared to VSL-only methods.

Our design provides two significant advantages in overcoming the aforementioned challenges. First, our regulation model requires only connectivity functionalities in vehicles, avoiding reliance on high levels of autonomy. Consequently, it regulates a wide range of vehicle types, including human-driven vehicles (HVs), by equipping them with vehicle-to-infrastructure communication technologies. Our proposed approach enables broader compatibility and achieves a higher regulation rate compared to CAV-specific control methods. Second, the macroscopic traffic flow representation substantially reduces data transmission requirements and eliminates the need for precise vehicle positioning, thereby enhancing scalability and practicality in real-world applications.

This paper theoretically addresses in-domain distributed control of PDE systems using a robust MARL framework. Two closely related works are \cite{belletti2017expert,yu2021reinforcement}, which propose scalable ramp metering control via MARL applied to the traffic PDE models. Several studies have also demonstrated the successful application of MARL to PDE dynamics in diverse domains, including fluid flow control, drag reduction, and turbulence modeling \cite{vignon2023effective,sonoda2023reinforcement,guastoni2023deep}, and  \cite{peitz2024distributed} further provides a theoretical foundation for distributed PDE control using MARL. To address the concern about the inconsistent improvements between theoretical macroscopic analysis and microscopic simulations \cite{zhang2016combined}, we adopt a practical evaluation strategy\textemdash directly testing in a microscopic simulator.

The main contributions of this work are as follows:
\begin{enumerate}
    \item We propose a novel fine-resolution, lane-grid-based lane-change regulation approach for freeway traffic control. The infrastructure-based framework is leveraged to enhance traffic efficiency by regulating lane-change maneuvers at a macroscopic level via a CTDE paradigm.
    \item In contrast to traditional MARL-based vehicle control methods, the proposed traffic flow control strategy simplifies communication requirements and reduces reliance on precise vehicle positioning, enabling efficient and scalable implementation in large-scale traffic systems.
    \item We establish a training framework by utilizing a microscopic simulator to generate microscopic vehicle trajectories and adopting the multi-lane macroscopic traffic PDEs to compute aggregated lane-changing regulation for CVs. The experimental results show improvements in traffic performance.
\end{enumerate}

To the best of our knowledge, this is the first attempt to employ CVs to regulate discretionary lane change at the lane-grid level. Positioned between infrastructure-level control and direct lane-change maneuver control, our regulation design minimizes information collection burdens while managing lane changes at a macroscopic level without introducing additional safety risks.

This paper is organized as follows. Section \ref{sec:methodology} details the modeling of our control problem using PDEs and derives the distributed regulation framework using multi-agent reinforcement learning. Section \ref{sec:sim_env_set} describes the setup of our experimental environments in the SUMO microscopic simulator. Section \ref{sec:result} presents extensive numerical results demonstrating the performance of the proposed model. Finally, Section \ref{sec:conclusion} concludes the paper.

\section{Lane Change Regulation Model} \label{sec:methodology}
In this section, we first describe the freeway traffic dynamics with lane-change using a generic multi-lane macroscopic traffic flow model, and then propose a multi-agent reinforcement learning control formulation to address this problem.
\subsection{Multi-lane Macroscopic Traffic Models}
While the first-order LWR model \cite{lighthill1955kinematic, richards1956shock} fails to capture stop-and-go waves due to its assumption of an equilibrium relationship between vehicle speed and density, second-order traffic models, such as the Aw-Rascle-Zhang (ARZ) model \cite{aw2000resurrection, zhang2002non}, introduce an additional velocity equation to account for traffic waves, acceleration, and deceleration effects. To model lane-change behavior in multi-lane settings, extended models treat each lane separately and couple them through lane-change dynamics.

We consider a highway of length $L$ with $m \in \mathbb{N}$ lanes, where lane 0 denotes the rightmost lane in the direction of travel. Traffic evolves over a time period $T$. For each lane $\alpha$, $\rho_{\alpha}(x,t)$ and $v_{\alpha}(x,t)$ denote the traffic density and average speed at position $x$ and time $t$, respectively. The general second-order model incorporating lane-change behavior is:
\begin{align}
    \begin{split}
        \frac{\partial \rho_{\alpha}}{\partial t} + \frac{\partial (\rho_{\alpha}v_{\alpha})}{\partial x} &=  S_1(\rho, v), \\
        \frac{\partial (\rho_{\alpha}v_{\alpha})}{\partial t} + \frac{\partial (\rho_{\alpha}v^2_{\alpha})}{\partial x} = \rho_{\alpha}&R(\rho_{\alpha}, v_{\alpha})-\frac{\partial p(\rho_{\alpha})}{\partial x} +   S_2(\rho, v),
    \end{split}
    \label{eq:general_one_lane}
\end{align}
where the relaxation term $R(\rho_{\alpha}, v_{\alpha})$ describes how vehicles adjust to their steady-state speed corresponding to the local density, and the pressure term $p(\rho_{\alpha})$ captures the response of the vehicle ensemble to density gradients \cite{treiber2013traffic}. The source terms $S_1(\rho, v)$ and $S_2(\rho, v)$ quantify the mass and momentum exchange due to lane changes, respectively. Here, $\rho=(\rho_0,\dots,\rho_{m-1})$ and $v=(v_0,\dots,v_{m-1})$ denote the density and speed states of all lanes. Following prior studies \cite{song2019second, yu2021output}, the source terms can be explicitly written as:
\begin{align}
    \begin{split}
        S_1(\rho, v) & = \frac{\rho_{\alpha-1}}{f_{\rho, \alpha-1}^{\text{left}}}a^{\text{left}}_{\alpha-1} + \frac{\rho_{\alpha+1}}{f_{\rho,\alpha+1}^{\text{right}}}a^{\text{right}}_{\alpha+1} \\
        & - \frac{\rho_{\alpha}}{f_{\rho, \alpha}^{\text{left}}}a^{\text{left}}_{\alpha}
        - \frac{\rho_{\alpha}}{f_{\rho,\alpha}^{\text{right}}}a^{\text{right}}_{\alpha},\\
        S_2(\rho, v) & = \frac{\rho_{\alpha-1}v_{\alpha-1}}{f_{\rho v, \alpha-1}^{\text{left}}}a^{\text{left}}_{\alpha-1} + \frac{\rho_{\alpha+1}v_{\alpha+1}}{f_{\rho v,\alpha+1}^{\text{right}}}a^{\text{right}}_{\alpha+1} \\
        &- \frac{\rho_{\alpha}v_{\alpha}}{f_{\rho v, \alpha}^{\text{left}}}a^{\text{left}}_{\alpha}- \frac{\rho_{\alpha}v_{\alpha}}{f_{\rho v, \alpha}^{\text{right}}}a^{\text{right}}_{\alpha},\\
    \end{split}
    \label{eq:general_source_control}
\end{align}
\noindent where $f^{\text{left, right}}_{\rho,\alpha}(\rho, v)$ and $f^{\text{left, right}}_{\rho v,\alpha}(\rho,v)$ are transition rate functions defined over space and time, representing vehicle mass and momentum exchange from lane $\alpha$ to its adjacent lanes. These functions can be modeled and calibrated based on vehicle interactions and traffic regulations~\cite{shvetsov1999macroscopic, porfyri2017calibration}.

The actions $a^{\text{left}}_{\alpha}(x,t), a^{\text{right}}_{\alpha}(x,t) \in \mathbb{R}$ regulate the magnitude of mass and momentum exchange due to lane changes in the respective directions. The uncontrolled case, where lane changes are not subject to regulation, corresponds to $a^{\text{left}}_{\alpha}(x,t) \equiv 1$ and $a^{\text{right}}_{\alpha}(x,t) \equiv 1$ for all $(x, t)$. This formulation allows the uncontrolled scenario to be viewed as a special case of the general model.

When modeling all lanes jointly, we define the global state $y = (\rho_0, v_0, \dots, \rho_{m-1}, v_{m-1})^T$ and control action $a = (a^{\text{left}}_0, a^{\text{right}}_0, \dots, a^{\text{left}}_{m-1}, a^{\text{right}}_{m-1})^T$. The system is described using the following partial differential operator:
\begin{align}
    \frac{\partial y}{\partial t} = \mathcal{N}(y, a) = -A(y)\frac{\partial y}{\partial x} + S(y, a),\label{eq:general_source_control_all_lane}
\end{align}
\noindent where $A(y)$ governs the characteristic speeds, and $S(y, a)$ represents the aggregated mass and momentum exchange across all lanes.

\subsection{Grid-based Lane Change Regulation Strategies} \label{sec:methodology_action}

We discretize the system both temporally and spatially. The spatial domain $[0, L]$ is divided into equally spaced intervals, each with grid length $\Delta x$, such that a lane is divided into $N_x = L/\Delta x$ grids. The time domain $[0, T]$ is partitioned with time-steps $\Delta T = t_q - t_{q-1}$, where $q = 1, \dots, N_T$.

Let {$\mathcal{V}(t_k)=\{1,2,\dots,n_{\mathrm{veh}}(t_k)\}$} denote the set of vehicles at time $t_k$, where {$n_{\mathrm{veh}}(t_k)$} is the number of vehicles in the system. For vehicle {$j\in\mathcal{V}(t_k)$}, let {$\ell^j(t_k)$} denote its lane index, and let {$x^j(t_k)$} and {$u^j(t_k)$} denote its longitudinal position and velocity, respectively. To incorporate the microscopic data into a PDE model, we need to compute the macroscopic traffic state for each spatial grid. \cite{fan2013data} employs a kernel density estimation (KDE) method to estimate density, velocity, and flow functions for the PDE model. Our approach adopts a simpler version of KDE, where each grid is represented by an averaged quantity, rather than requiring detailed positional data, which reduces the need for precise vehicle positioning.

Specifically, the density at the $i$-th grid in lane $\alpha$ is computed by counting the number of vehicles within the grid at $x_i$, and the velocity is estimated by averaging their speeds:
\begin{align}
    \begin{split}
    N_{\alpha_i}(t_k) &= {\sum_{j\in\mathcal{V}}}
    \mathbf{1}_{[x_i-\frac{\Delta x}{2},x_i+\frac{\Delta x}{2}],~{\ell^j=\alpha}}
    ({x^j}),\\
    \rho_{\alpha_i}(t_k) &= \frac{N_{\alpha_i}(t_k)}{\Delta x}, \\
    v_{\alpha_i}(t_k) &= \frac{1}{N_{\alpha_i}(t_k)}
    {\sum_{j:x^j\in[x_i-\frac{\Delta x}{2},x_i+\frac{\Delta x}{2}],~\ell^j=\alpha}}
    {u^j}.
    \end{split}
    \label{eq:grid_density_velocity_cal}
    \end{align}
\noindent In practice, average speed and density in each grid are acquired through measurements from local detectors, or reports from CVs \cite{belletti2017expert,bekiaris2020pde}. The flow is then calculated by multiplying the estimated density and velocity:
        $ q_{\alpha_i}(t_k) = \rho_{\alpha_i}(t_k) \cdot v_{\alpha_i}(t_k).$

Within our system, we assume the presence of two types of vehicles: HVs, which execute lane changes without intervention from the regulation model, and CVs, which are HVs equipped with connectivity technologies that adhere strictly to regulation signals. Notably, both CVs and HVs follow the same microscopic dynamic models.
In the discretized setting of Eq.~(\ref{eq:general_source_control}), our model selectively controls lane changes of CVs by permitting or prohibiting transitions from grid $\alpha_i$ to its adjacent lane grids.
\begin{figure}[t]
\centering
\includegraphics[width=3.6in]{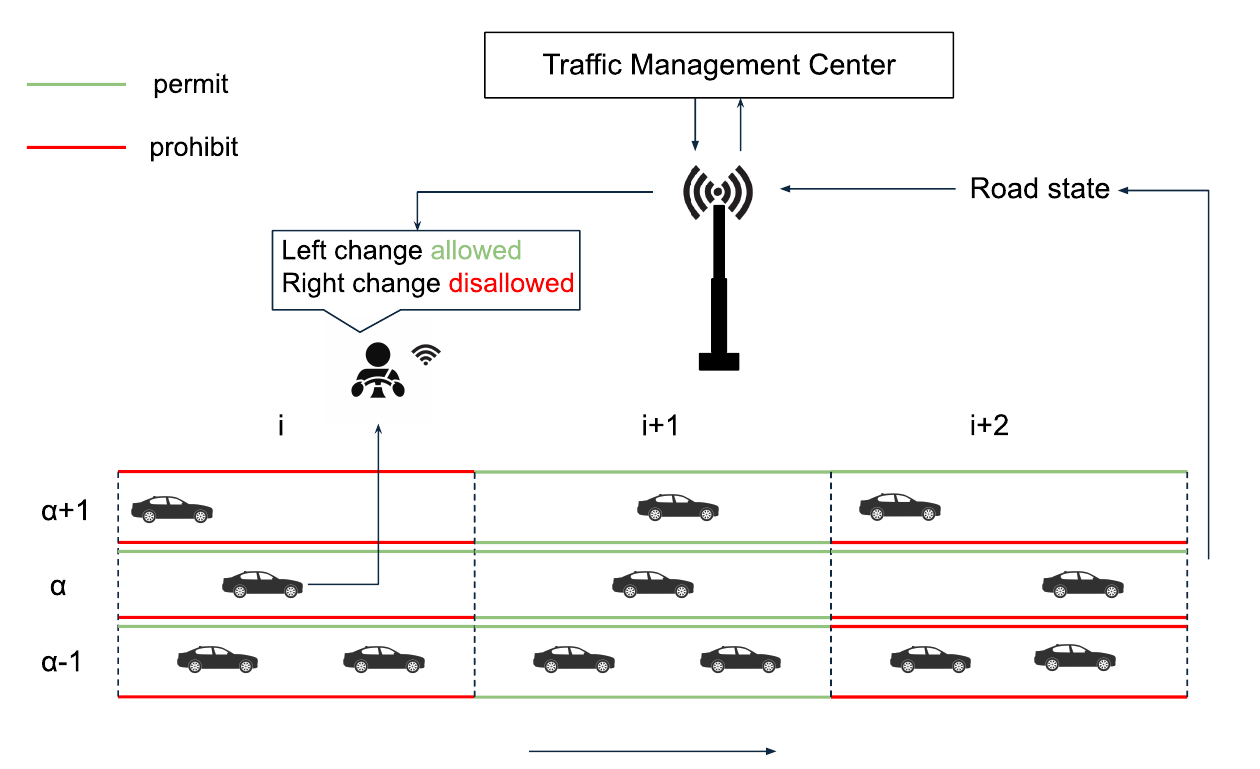}
\caption{Model demonstration of lane change regulation signals broadcast to drivers via traffic management center.}
\label{fig:method_demo}
\end{figure}
Specifically, the action for any grid $\alpha_i$ at time $t_k$ is discretized as a two-dimensional vector:
\begin{align}
    a_{\alpha_i}(t_k) = (a_{\alpha_i}^{\text{left}}(t_k), a_{\alpha_i}^{\text{right}}(t_k)) \in \{0, 1\}^2.
    \label{eq:e007}
\end{align}
\noindent Each component indicates whether lane change requests from grid $\alpha_i$ in a specific direction are permitted at time $t_k$. If a lane change is disallowed, the corresponding gain or loss term in Eq.~(\ref{eq:general_source_control}) is suppressed by multiplying with zero. Otherwise, the source terms behave as in the uncontrolled case, where all actions are set to one. Fig.~\ref{fig:method_demo} illustrates how our model functions in a real-world scenario by broadcasting lane change regulations to vehicles through road infrastructure.

In the context of reinforcement learning, the discretized transition function expressed using the partial differential operator in Eq. (\ref{eq:general_source_control_all_lane}) for each grid column is:
\begin{align}
    \begin{split}
        y_{x=x_i}(t_{k+1}) &= g_{x_i}(y(t_{k}), a(t_{k})) \\
        &= y_{x=x_i}(t_{k}) + {\int_{t_k}^{t_{k+1}}\mathcal{N}(y(\tau), a(t_{k}))d\tau}|_{x=x_i},
    \end{split}
    \label{eq:transition_function}
\end{align}

\noindent where the action $a_{x_i}(t_k)$ is held constant during the time interval (zero-order hold).

Two important properties of traffic PDE models simplify the formulation of the transition function. First, the hyperbolic structure ensures finite wave propagation speeds, so the evolution of a grid over a short time depends only on a localized region of the whole traffic state. Second, the dynamics in Eq. (\ref{eq:general_source_control_all_lane}) exhibit spatial translation invariance, meaning the system's behavior does not depend on specific spatial locations when border effects are negligible. As a result, the transition function for all grid columns in Eq. (\ref{eq:transition_function}) can be uniformly approximated by a shared function $\tilde{g}$, enabling parameter sharing across all grid columns.

Two further approximations are made in the training algorithm design:

\begin{enumerate}
    \item We assume identical geometric characteristics, speed limits, and lane discipline across all lanes in each simulated environment.
    Neglecting border effects (e.g., where border lanes have only one adjacent lane), spatial translation invariance extends to all grids. Consequently, we assign one agent to each grid and train a model with shared parameters across the entire grid system.
    \item The dynamics during safety-critical events may evolve with a spatially inhomogeneous partial differential operator, deviating from Eq. (\ref{eq:general_source_control_all_lane}). However, these anomalies are confined to a small fraction of the overall space-time domain and thus be neglected in experiment. 
\end{enumerate}

\subsection{Multi-agent Reinforcement Learning}

\begin{figure*}[t]
\centering
\includegraphics[width=7in]{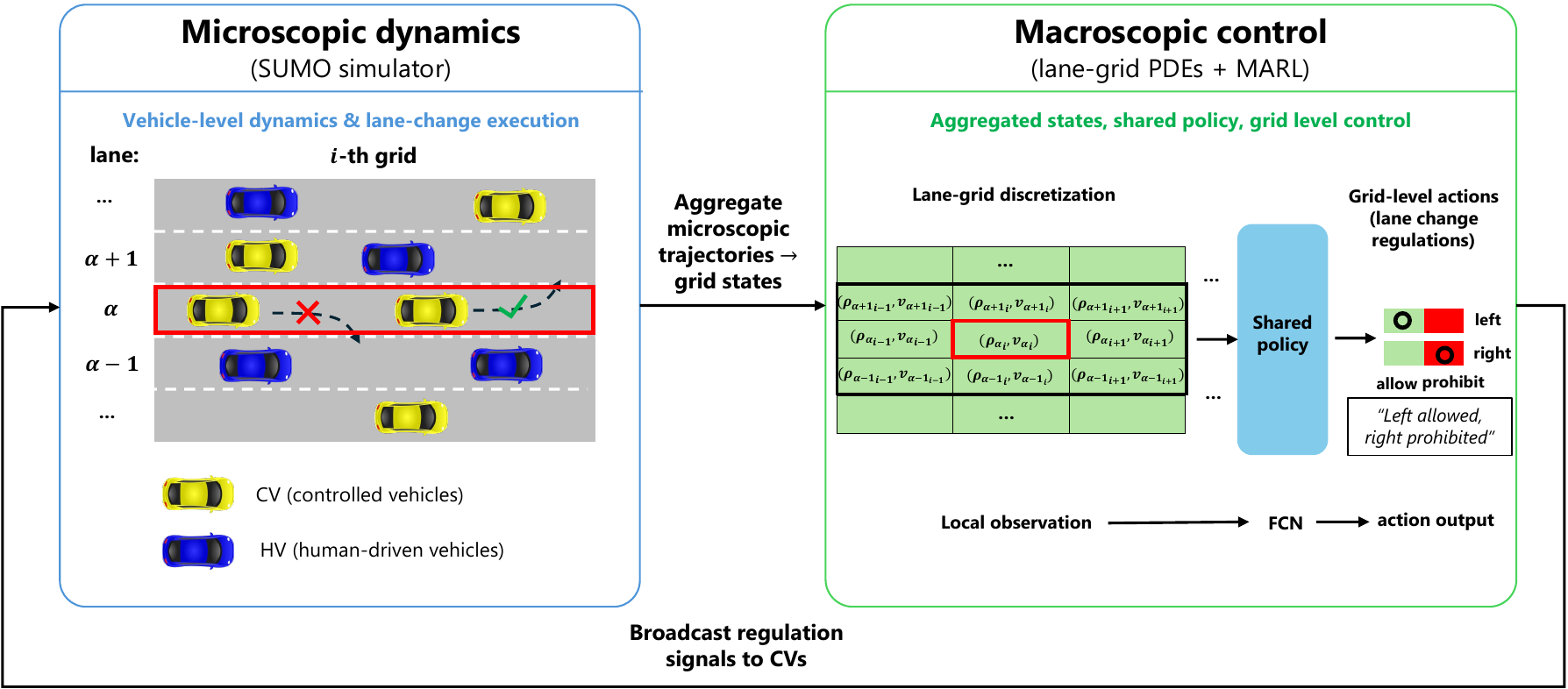}
\caption{The architecture of the proposed lane-change regulation model, integrating the microscopic traffic simulator SUMO for vehicle-level dynamics and macroscopic traffic PDEs for aggregated lane-change decisions.}
\label{fig:overall_framework}
\end{figure*}

We address the control problem using multi-agent reinforcement learning. The overall architecture of the proposed lane change regulation system is illustrated in Fig.~\ref{fig:overall_framework}. Each grid has an agent, and all agents share network parameters. The target system is formulated as a Partially Observable Markov Decision Process (POMDP). Formally, a POMDP can be represented as a 6-tuple $(\mathcal{S}, \mathcal{A}, \mathcal{P}, \mathcal{R}, \mathbf{\Omega}, \mathcal{O})$, where $\mathcal{S}$ denotes the state space, $\mathbf{\Omega}$ represents the observation space, $\mathcal{A}$ is the action space, $\mathcal{P}:\mathcal{S} \times \mathcal{A} \rightarrow \mathcal{S}$ describes the state transition distribution, $\mathcal{O}$ is the probabilistic observation model, and $\mathcal{R}$ is the reward function.

\textit{1) State space:} The entire state space of the highway is defined as { $\mathcal{S}=\prod_{\alpha=0}^{m-1}\prod_{i=1}^{N_x}S_{\alpha_i},$ where the product symbol denotes the Cartesian product over all lane-grid state spaces.} Here $S_{\alpha_i}$ represents the state of the grid $\alpha_i$. Features encoded in each state with the grid include:
\begin{itemize}
    \item $\rho$: the total vehicle density.
    \item $\rho_c$: the CV density, where $\rho_c \in [0, \rho] $. Here, $\rho_c = \rho$ indicates that all vehicles in the grid are {connected vehicles and are subject to regulation}.
    \item $v$: the average longitudinal vehicle velocity.
    \item $v_c$: the average longitudinal velocity of CVs.
\end{itemize}
\noindent In Section \ref{sec:methodology_action}, we derive the dynamics model in a fully {connected environment where all vehicles are subject to regulation.} This model can be extended to a partially {connected} environment where both HVs and CVs coexist. We assume that HVs and CVs follow the same microscopic models, as CVs are essentially HVs equipped with connectivity functionalities. This assumption of driving behavioral homogeneity ensures that the fundamental diagram and the transition rates $f_{\rho}$ and $f_{\rho v}$ in Eq. (\ref{eq:general_source_control}) depend only on the combined density, velocity, and flow of both CVs and HVs. The evolution of density and velocity for either CV or HV is governed by the same equations, as expressed in Eq. (\ref{eq:general_source_control}), while regulation actions are applied only to CVs.

The observation space for each grid agent $\mathbf{\Omega}_{\alpha_i}$ consists of a localized subset of the global state, defined as $\mathbf{\Omega}_{\alpha_i} = \{ S_{\mathcal{I}_{\alpha}, \mathcal{N}_i} \}$, where $\mathcal{N}_i \subset \{1, \dots, N_x\}$ represents the set of neighboring grids along the longitudinal direction, and $\mathcal{I}_{\alpha} \subset \{0, \dots, m-1\}$ denotes the set of adjacent lanes. In practice, the scope of $\mathcal{N}_i$ and $\mathcal{I}_{\alpha}$ is determined by the receptive field of the convolutional layers used in the policy and value networks.

\textit{2) Action space:} As described in Section \ref{sec:methodology_action}, the action space $\mathcal{A}_{\alpha_i}$ for the grid $\alpha_i$ is two-dimensional. Only the lane change requests from CVs are regulated by the action output.
\begin{align}
    a_{t, \alpha_i} \sim \pi_{\theta}(\mathbf{o}_{t,\alpha_i})
    \label{eq:e014}
\end{align}
\noindent It is important to note that lane change requests from individual vehicles already meet the safety criterion defined in the microscopic physical models. This facilitates the design of the reward function for the proposed model $\pi_{\theta}$, as safety measures do not need to be explicitly included.

\textit{3) Reward function:} The reward function evaluates the local performance of each grid agent, focusing primarily on traffic efficiency by assessing congestion levels. A comprehensive review of road traffic congestion measures is provided in \cite{afrin2020survey}.

The reward $r_1$ measures efficiency based on the speed performance index \cite{he2016traffic}. For a given grid agent $\alpha_i$ with local area $\mathcal{B}(\alpha_i)$, $r_1$ is defined as:
\begin{align}
    r_1 = \frac{1}{{|\mathcal{B}(\alpha_i)|}} \sum_{\tilde{\alpha}_k\in\mathcal{B}(\alpha_i)} \frac{v_{\tilde{\alpha}_k}}{v_{max}}.
    \label{eq:e12}
\end{align}

\noindent Here, the aggregated speed information of all vehicles in a grid is evaluated, rather than just {connected} vehicles. This approach is chosen because the objective is to improve overall traffic efficiency through the actions of CVs.

We also consider traffic density by defining $r_2$ based on the volume-to-capacity ratio (V/C):
\begin{align}
    r_2 = \frac{1}{{|\mathcal{B}(\alpha_i)|}} \sum_{\tilde{\alpha}_k\in\mathcal{B}(\alpha_i)} \left( 1 - \frac{\rho_{\tilde{\alpha}_k}}{\rho_{max}} \right).
    \label{eq:e14}
\end{align}

\noindent Unlike the discretized Level of Service (LOS) indicator used in \cite{afrin2020survey}, we adopt a continuous function with respect to $\rho$ as defined in Eq. (\ref{eq:e14}). This choice ensures training stability, especially in scenarios with high-frequency oscillations, whereas the LOS value can fluctuate significantly over time.  In Eq. (\ref{eq:e12}) and (\ref{eq:e14}), $v_{max}$ and $\rho_{max}$ are determined by the simulation settings, as shown in Table \ref{tab:train_params}. The $\mathcal{B}(\alpha_i)$ used in all the experiments includes all grids within a rectangular area:
$\mathcal{B}(\alpha_i) = \{\tilde{\alpha}_k|\tilde{\alpha}=\alpha-1,\alpha,\alpha+1; |x_i-x_k|\leq 2\Delta x\}.$
The overall reward function for grid agent $\alpha_i$ is defined as a weighted sum of $r_1$ and $r_2$:
\begin{align}
    r = w_1 r_1 + w_2 r_2.
    \label{eq:e15}
\end{align}

\section{Experimental setup}
\begin{table}[t]
\caption{Primary hyperparameters used during training.\label{tab:train_params}}
\centering
\renewcommand{\arraystretch}{1.3} 
\begin{tabular}{cc}
\hline
\hline
Parameters & Value \\
\hline
lane grid length (m) & 100 \\
discount factor $\gamma$ & 0.95 \\
reward weight of speed $\omega_1$ & 0.5 \\
reward weight of density $\omega_2$ & 0.5 \\
road max density (veh/(lane$*$m)) & 0.133  \\
road max speed (m/s) & 24.59 \\
\hline
\hline
\end{tabular}
\end{table}
\subsection{Simulation Environment and Settings} \label{sec:sim_env_set}
\begin{table}[t]
\caption{Parameter setting for IDM model\label{tab:IDM_params}}
\centering
\renewcommand{\arraystretch}{1.3} 
\begin{tabular}{ccc}
\hline
\hline
Symbol & Description & Value \\
\hline
$L$ & Vehicle length & 5 m \\
$v_0$ & Desired speed & 24.59 m/s \\
$T$ & Time gap & 1.4 s \\
$s_0$ & Minimum gap & 2.5 m \\
$\delta$ & Acceleration exponent & 4 \\
$a$ & Acceleration & 0.73 $m/s^2$ \\ 
$b$ & Comfortable deceleration & 1.67 $m/s^2$ \\
\hline
\hline
\end{tabular}
\end{table}

\begin{figure}[t]
\centering
\includegraphics[width=2.3in]{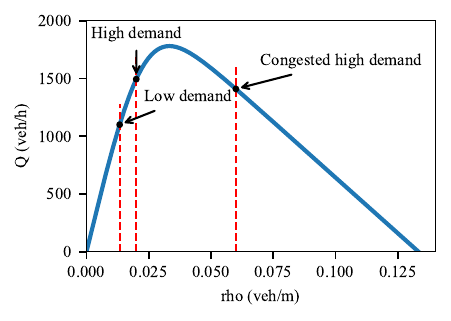}
\caption{Demand settings with the fundamental diagram.}
\label{fig:idm_demand}
\end{figure}

We build a 5-lane highway road in the SUMO microscopic simulator \cite{lopez2018microscopic}. The road length is 1000 m and lane width is 3.2 m. The speed limit of each lane is set to 24.59 m/s (55 mph). Vehicles arrive from the entrance of the road, and the number of vehicles emitted each second follows a binomial distribution. The intelligent driver model (IDM) \cite{treiber2000congested} is selected as the car-following model for all simulated vehicles. The corresponding parameter settings are shown in Table \ref{tab:IDM_params}. In addition, we set the lateral resolution to 0.8m and choose the SL2015 model \cite{erdmann2015sumo} as the lane-changing model. Table \ref{tab:IDM_fd} shows the traffic demands in the simulation scenarios. The relation of inflow and density is decided by the fundamental diagram derived from IDM car-following model, as illustrated in Fig. \ref{fig:idm_demand}. 

\begin{table}[t]
\caption{Traffic demand settings in simulation scenarios\label{tab:IDM_fd}}
\centering
\renewcommand{\arraystretch}{1.3} 
\begin{tabular}{cccc}
\hline
\hline
 & Inflow   & Density & Velocity  \\
 & (veh/(lane*h)) & (veh/(lane*m)) & (m/s) \\
\hline
Low demand & 1100 & 0.013 & 22.93  \\
High demand & 1495 & 0.02 & 20.76  \\
Congested High demand & 1410 & 0.06 & 6.53  \\
\hline
\hline
\end{tabular}
\end{table}

\begin{algorithm}[t]
\caption{The Pseudo-code for training the regulation policy based on Double DQN}
\begin{algorithmic}
    \STATE \textbf{Initialize:} Q-network $Q_{\theta}$, target network $Q_{\theta^-}$, and replay buffer $\mathcal{D}$.
    \STATE \textbf{For} step=1, ..., M
    \STATE \hspace{0.5cm} Reset the environment state $s_0$.
    \STATE \hspace{0.5cm} \textbf{While} not \textit{done}
    \STATE \hspace{1cm} \textbf{For each agent}
    \STATE \hspace{1.5cm} Select action with exploration probability $\epsilon$
    \STATE \hspace{3cm} $a_{t, \alpha_i} = \arg\max_a Q_{\theta}(\mathbf{s}_{t,\alpha_i}, a)$
    \STATE \hspace{1cm} \textbf{End}
    \STATE \hspace{1cm} {Execute $a_t$ in SUMO and observe the aggregated lane-grid state $s_{t+1}$.}
    \STATE \hspace{1cm} Store global transition in replay buffer $\mathcal{D}$
    \STATE \hspace{3cm} $\mathcal{D}\leftarrow\mathcal{D}\cup (\mathbf{s}_t, a_t, r_t, \mathbf{s}_{t+1})$
    \STATE \hspace{1cm} Sample $B=\{(\mathbf{s}_k, a_k, r_k, \mathbf{s}_{k+1})\}$ from $\mathcal{D}$
    \STATE \hspace{1cm} \textbf{For each agent}
    \STATE \hspace{1.5cm} Calculate agent reward  $r_{j,\alpha_i}$ by using Eq.~(\ref{eq:e15}).
    \STATE \hspace{1.5cm} Compute target Q value:
    \STATE \hspace{2.0cm} $y_{j,\alpha_i} = r_{j,\alpha_i} +$
    \STATE \hspace{2.7cm} $\gamma Q_{\theta^-}(\mathbf{s}_{j+1,\alpha_i}, \arg\max_a Q_{\theta}(\mathbf{s}_{j+1,\alpha_i}, a))$
    \STATE \hspace{1cm} \textbf{End}
    \STATE \hspace{1cm} Reshape batch into $\tilde{B}$ with size of $|B|\times num_{agents}$.
    \STATE \hspace{1cm} Perform gradient descent on loss
    \STATE \hspace{3cm} $\mathcal{L}(\theta) = \mathbb{E}_{\tilde{B}}[(y_j - Q_{\theta}(\mathbf{s}_j, a_j))^2]$
    \STATE \hspace{1cm} \textbf{If} step mod $T_{update}$ == 0
    \STATE \hspace{1.5cm} Update target network $Q_{\theta^-} \leftarrow Q_{\theta}$.
    \STATE \hspace{1cm} \textbf{End}
    \STATE \hspace{1cm} \text{step} $\leftarrow$ \text{step+1}
    \STATE \hspace{0.5cm} \textbf{End}
    \STATE \textbf{End}
\end{algorithmic}
\label{alg:1}
\end{algorithm}

Given a demand setting, we test our model in different scenarios.
In addition to a stable inflow scenario, typical safety-critical events are included to evaluate the model's performance under transient and hazardous traffic conditions.
{The scenarios are described as follows.}
\begin{enumerate}
    \item \textit{Stable Flow}: In a stable flow scenario, all lanes share a constant inflow rate selected from Table \ref{tab:IDM_fd}. All vehicles depart from the entry position of the road.
    \item \textit{Lane Degrade}: In a lane degrade scenario, {one or more lanes can experience reduced capacity} due to narrow lane width, dangerous road conditions, or reduced visibility range. This can be simulated by assigning an increased time gap parameter in the IDM model to vehicles in the affected lane\cite{treiber2000congested}. Built upon a \textit{Stable Flow} scenario, we create lane degrade events with random lane, location range, and time range. The new time gap is set in the range $[4.0, 10.0]$ seconds.
    Lane change actions are prohibited for all vehicles traveling in hazardous lane grids, regardless of the decisions made by the driver or our model.
    \item \textit{Vehicle Stop}: In a vehicle stop scenario, a lane block event is created by an accidental vehicle. We simulate this event in this way: at a given trigger time, a vehicle will decelerate relatively aggressively until the speed reaches zero. Then the vehicle remains stopped for a random period of time. During the recovery stage, the accidental vehicle and its blocked followers will resume their route. 
\end{enumerate}

Scenario identification is essential for selecting the appropriate policy for deployment. In practice, real-time road conditions can be inferred from official sources (e.g., transportation agencies) and crowdsourced GPS data. Commercial navigation apps like Google Maps use such data to detect incidents and congestion \cite{d2017detection}, which can also support timely policy switching in our system.
\subsection{Training}
We train our policy using the Double Deep Q-Network (Double DQN) algorithm, which is well-suited for tasks with discrete action spaces. Double DQN addresses the overestimation bias commonly found in traditional Deep Q-Networks (DQN) by decoupling action selection from value estimation \cite{van2016deep}. This improves training stability and results in more accurate value function estimates. The full training procedure is detailed in Algorithm \ref{alg:1}. To evaluate the effectiveness of our method, we compare it against a benchmark policy trained using the Proximal Policy Optimization (PPO) algorithm, a widely used on-policy reinforcement learning method known for its sample efficiency and robustness.

For each scenario and demand condition, we train a separate policy network which employs a fully convolutional neural network (FCN) with ReLU activation functions and a discount factor of 0.95. A training environment is created using the SUMO. Simulation parameters include a unified simulation step $\delta=0.1s$, a reward step $\Delta t_r=1.0s$, and an environmental step size $\Delta t_{env}=4.0s$ for consistency across environments. The reward for each step is the average reward calculated over the interval $\Delta t_{env}$, with samples taken at regular intervals of $\Delta t_r$. The lane grid length $\Delta x_{env}$ is set to 100\,m, as summarized in Table~\ref{tab:train_params}. At the end of each episode, the training environment is reset by generating the traffic demand using a random seed. In scenarios involving safety-critical events, additional parameters such as the location range and affected lanes are also randomized to improve generalization.

Training is conducted over 250{,}000 steps, with performance metrics calculated at the end of each episode. {The experimental environment is implemented using SUMO, the Traffic Control Interface (TraCI), and OpenAI Gym APIs. All experiments are conducted on a computer equipped with an Nvidia GeForce RTX 4090 GPU and 64~GB RAM. The software environment uses Python 3.9 and PyTorch 1.13.1.}

\subsection{Evaluation}
The primary goal of the proposed model is to optimize the efficiency of the entire freeway section. At the same time, we need to monitor whether our model causes negative effects on driving safety and energy cost. Therefore, traffic efficiency, safety, and energy cost are the performance metrics of interest. In this context, we use average travel speed of all vehicles
for evaluating traffic efficiency. {Energy cost is measured by CO$_2$ emission.} We adopt the {\textit{HBEFA4/PC\_petrol\_Euro-6d}} emission model for all simulated vehicles. Since our model does not intervene directly on vehicle-level maneuvers, {vehicle-level collision avoidance is handled by the microscopic simulator, and no collisions are observed in our evaluations.} Besides the collision rate indicator, time-to-collision (TTC) based measures are also commonly used \cite{monteiro2023safe}, \cite{behbahani2015new}. When TTC is less than a threshold, a collision risk is indicated. We use two TTC-based indicators: average TTC and time exposed TTC (TET) \cite{minderhoud2001extended}. Compared to average TTC, time exposed TTC is normalized by travel time of each vehicle. {Definitions of the} two safety indicators are given in Eq. (\ref{eq:e1}) and Eq. (\ref{eq:e2}). {The critical threshold TTC$^*$ is selected according to the traffic condition. For the low and high demand scenarios, we use TTC$^*=1.5$s, following common practice in surrogate safety analysis \cite{virdi2019safety,gettman2008surrogate}. For the congested high demand scenario, we use TTC$^*=5.0$s as a more conservative threshold to capture potential conflicts earlier under low-speed and small-gap traffic conditions \cite{minderhoud2001extended}.}
\begin{align}  \overline{\textnormal{TTC}} = \frac{1}{N_{veh}}\sum_{i=1}^{N_{veh}}\sum_{t=1}^{T_i}\mathbf{1}_i(TTC_i(t)<TTC^*) \cdot\delta,
    \label{eq:e1}
\end{align}
\begin{align}
    \textnormal{TET} &= \frac{1}{N_{veh}}\sum_{i=1}^{N_{veh}}\sum_{t=1}^{T_i}\frac{1}{T_i}\cdot\mathbf{1}_i(TTC_i(t)<TTC^*).
    \label{eq:e2}
\end{align}
We evaluate models trained in each environment at varying checkpoints. The baseline model is defined to follow the SUMO IDM car-following model and SL2015 lane-changing model for all vehicles.
Since each traffic environment in SUMO is initialized with an empty road, all data collected during the warm-up phase is excluded when computing metrics in both the training and evaluation stages.
\section{Numerical Results} \label{sec:result}
This section evaluates the performance of the lane-change regulation model in different traffic scenarios and under different penetration rates of CVs.
We first evaluate the model's performance under a 100\% CV penetration rate, and then examine how it varies in mixed-traffic scenarios with partial CV penetration in Section \ref{sec:result_diff_rate}.
\subsection{Performance of the Lane Change Regulation Model} \label{sec:result_performance}
\begin{figure*}[t]
\centering
\vspace{-1em}
\subfloat{\includegraphics[width=2in]{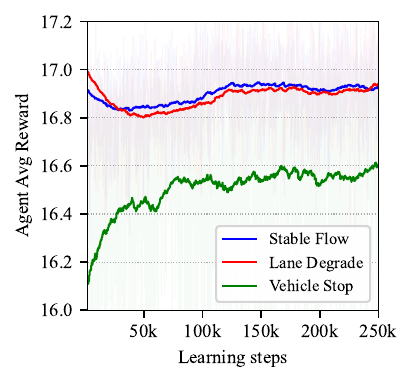}
}
\subfloat{\includegraphics[width=2in]{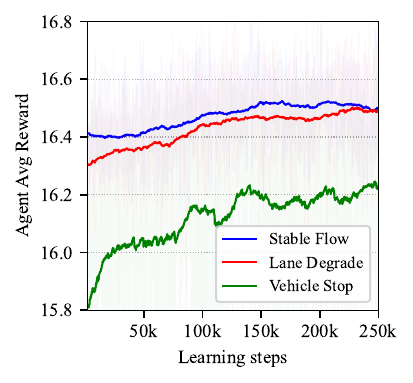}
}
\subfloat{\includegraphics[width=2in]{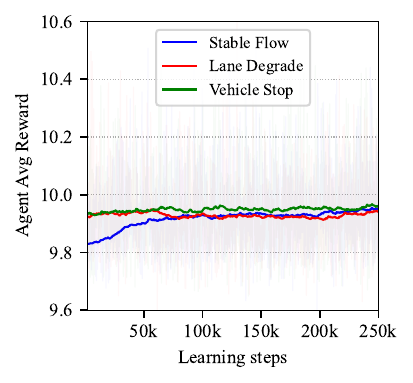}
} \\
\parbox[c]{\textwidth}{\centering \footnotesize Total Reward} \\
\vspace{-1em}
\subfloat{\includegraphics[width=2in]{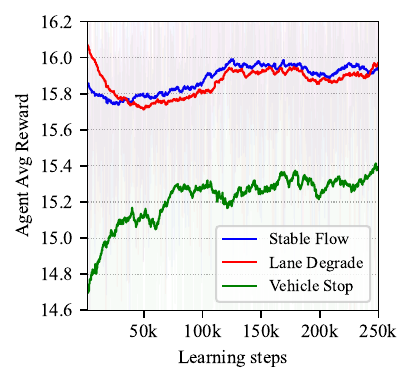}
}
\subfloat{\includegraphics[width=2in]{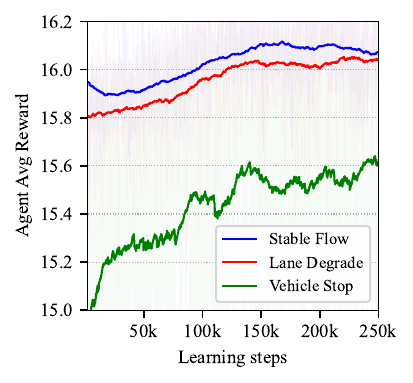}
}
\subfloat{\includegraphics[width=2in]{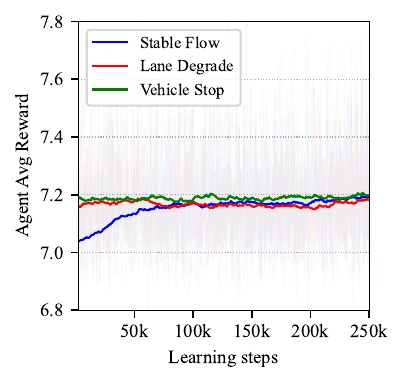}
} \\
\parbox[c]{\textwidth}{\centering \footnotesize Reward Term $r_1$} \\
\vspace{-1em}
\subfloat{\includegraphics[width=2in]{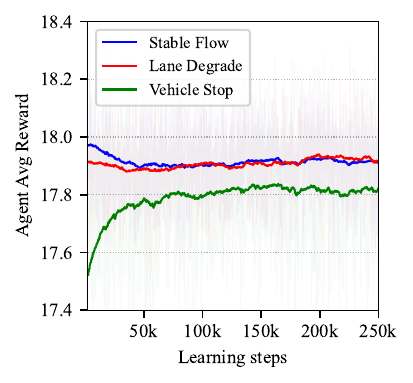}
}
\subfloat{\includegraphics[width=2in]{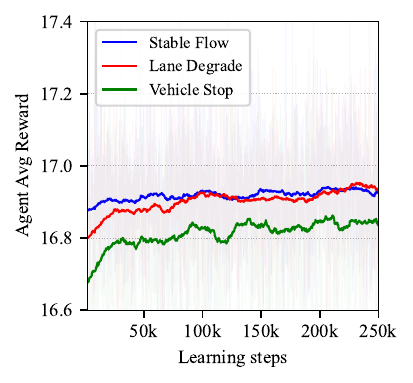}
}
\subfloat{\includegraphics[width=2in]{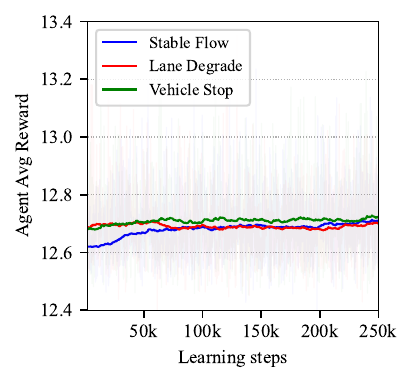}
}\\
\parbox[c]{\textwidth}{\centering \footnotesize Reward Term $r_2$} \\
\caption{Average agent reward per episode during the training phase under varying environments. From left to right, the columns represent low, high, and congested high demand settings. The top row shows the total reward, while the second and third rows display the rewards for terms $r_1$ and $r_2$, respectively. Solid lines represent an exponential moving average for smoothing.}
\label{fig:train_reward}
\end{figure*}

To minimize the dependency on precise parameter initialization and enhance training stability, the states are normalized to the range $[0, 1]$ by dividing each dimension by its respective maximum value. These maximum values of vehicle speed and density are determined by the fundamental diagram. The primary hyperparameters of our algorithm are listed in Table \ref{tab:train_params}. 
Fig. \ref{fig:train_reward} illustrates the average accumulated reward across agents per episode during the training phase. Under free-flow demand conditions (low and high), a vehicle stop event substantially reduces reward performance. In contrast, compared to congested high demand conditions, a significant improvement trend is observed under free-flow demand conditions across all scenarios. As shown in Fig. \ref{fig:train_reward}, this improvement is primarily attributed to the enhancement in the speed performance index reward term associated with vehicle travel speeds. This indicates that the policy seeks to optimize traffic speed without significantly changing the equilibrium state of the traffic, which is governed by the spatial distribution of traffic density.    

Table \ref{tab:performance_all} summarizes the results of our model and the baselines across all evaluation environments. Comparative metrics against the baselines are presented in Table \ref{tab:performance_uplift_all}, with arrows indicating the optimization direction for each indicator. Our model outperforms the baseline in terms of traffic efficiency under low and high demand conditions, while demonstrating comparable performance under congested high demand conditions. As shown in Table \ref{tab:train_lane_change_count}, the average number of lane changes per vehicle in the \textit{Stable Flow} scenario decreases {sharply} under congested high demand conditions compared to low demand conditions. {Specifically, it decreases by 87.1\% for the baseline and 89.5\% for our model.} Moreover, our regulation model suppresses lane-changing behavior compared to the baseline. For example, in the \textit{Stable Flow} scenario, the lane change frequency decreases by {7.5\%, 39.0\%}, and 25.0\% under low, high, and congested high demand conditions, respectively. This indicates a link between the optimization strategy of our model and the  suppression of lane change frequency. Consequently, it can be inferred that low lane change frequency in congested conditions restricts the optimization potential of our model, leading to its comparable performance under congested high demand conditions.

\begin{table*}[t]
  \caption{Performance metrics (mean and standard deviation) across scenarios under different demand conditions. \label{tab:performance_all}}
  \centering
  \renewcommand{\arraystretch}{1.3} 
  \begin{tabular}{p{0.8in}p{0.75in}p{0.7in}p{0.7in}p{0.7in}p{0.7in}p{0.7in}p{0.7in}}
  \hline
  \hline
  \multirow{2}{*}{\textbf{Metric}}        & \multirow{2}{*}{\textbf{Demand}} & \multicolumn{2}{c}{\textbf{Stable Flow}} & \multicolumn{2}{c}{\textbf{Lane Degrade}} &
  \multicolumn{2}{c}{\textbf{Vehicle Stop}} \\ \cmidrule(lr){3-4} \cmidrule(lr){5-6} \cmidrule(lr){7-8}
                                 &                         & Baseline   & Our model   & Baseline  & Our model   & Baseline  & Our model   \\ \hline
  \multirow{3}{*}{Average speed (m/s)} & Low  & 21.78 $\pm$ 0.12  & 22.38 $\pm$ 0.10 & 20.69 $\pm$ 0.94   & 21.60 $\pm$ 0.62  & 20.08 $\pm$ 1.30  & 21.11 $\pm$ 0.83    \\
  									& High & 19.29 $\pm$ 0.18 & 19.78 $\pm$ 0.17 & 18.46 $\pm$ 0.88 & 19.28 $\pm$ 0.52 & 18.07 $\pm$ 0.87 & 18.93 $\pm$ 0.61  \\
  									& Congested High & 8.54 $\pm$ 0.12  & 8.56 $\pm$ 0.15 & 8.87 $\pm$ 0.36  & 8.83 $\pm$ 0.33  & 8.84 $\pm$ 0.42  & 8.9 $\pm$ 0.47    \\ \hline
  \multirow{3}{*}{CO$_2$ emission (g)} & Low  & 122.68 $\pm$ 0.89  & 126.69 $\pm$ 0.70 & 126.12 $\pm$ 3.53   & 130.11 $\pm$ 3.13  & 132.76 $\pm$ 9.69  & 134.14 $\pm$ 6.58    \\
  									& High & 119.61 $\pm$ 1.06 & 122.73 $\pm$ 1.08 & 125.71 $\pm$ 5.77 & 128.14 $\pm$ 4.59 & 133.23 $\pm$ 11.27 & 132.89 $\pm$ 7.32  \\
  									& Congested High & 256.5 $\pm$ 3.24  & 256.25 $\pm$ 3.83 & 259.56 $\pm$ 11.85  & 259.93 $\pm$ 11.83  & 259.24 $\pm$ 7.83  & 258.87 $\pm$ 8.19    \\ \hline
  \multirow{3}{*}{$\overline{\textnormal{TTC}}$ (s)} & Low  & 0.00 $\pm$ 0.00  & 0.00 $\pm$ 0.00 & 0.04 $\pm$ 0.07   & 0.03 $\pm$ 0.07  & 0.05 $\pm$ 0.10  & 0.01 $\pm$ 0.02
  \\
  									& High & 0.04 $\pm$ 0.03 & 0.00 $\pm$ 0.01 & 0.16 $\pm$ 0.19 & 0.02 $\pm$ 0.03 & 0.13 $\pm$ 0.14 & 0.04 $\pm$ 0.08  \\
  									& Congested High & 3.01 $\pm$ 0.28  & 2.98 $\pm$ 0.29 & 2.86 $\pm$ 0.46  & 2.88 $\pm$ 0.46  & 2.7 $\pm$ 0.58  & 2.67 $\pm$ 0.62    \\ \hline
  \multirow{3}{*}{TET (\%)} & Low  & 0.00 $\pm$ 0.01  & 0.00 $\pm$ 0.00 & 0.05 $\pm$ 0.08   & 0.04 $\pm$ 0.09  & 0.07 $\pm$ 0.12  & 0.01 $\pm$ 0.02    \\
  									& High & 0.08 $\pm$ 0.05 & 0.01 $\pm$ 0.01 & 0.22 $\pm$ 0.20 & 0.03 $\pm$ 0.04 & 0.19 $\pm$ 0.16 & 0.06 $\pm$ 0.09  \\
  									& Congested High & 2.37 $\pm$ 0.23  & 2.34 $\pm$ 0.23 & 2.15 $\pm$ 0.29  & 2.18 $\pm$ 0.28  & 2.07 $\pm$ 0.46  & 2.05 $\pm$ 0.49    \\
  \hline
  \hline
  \end{tabular}
  \end{table*}

  \begin{table*}[h]
  \caption{Comparison of performance metrics between the proposed model and baselines across scenarios under varying demand conditions.\label{tab:performance_uplift_all}}
  \centering
  \renewcommand{\arraystretch}{1.3} 
  \begin{tabular}{p{0.85in}p{0.3in}p{0.3in}p{0.4in}p{0.3in}p{0.3in}p{0.4in}p{0.3in}p{0.3in}p{0.4in}}
  \hline
  \hline
  \multirow{2}{*}{\textbf{Metric}} & \multicolumn{3}{c}{\textbf{Stable Flow}} & \multicolumn{3}{c}{\textbf{Lane Degrade}} & \multicolumn{3}{c}{\textbf{Vehicle Stop}} \\
  \cmidrule(lr){2-4} \cmidrule(lr){5-7} \cmidrule(lr){8-10}
                                    & Low  & High  & Congested High  & Low  & High  & Congested High  & Low  & High  & Congested High  \\ \hline
  Average Speed ($\uparrow$)     & 2.8  & 2.6  & 0.3   & \textbf{4.4}  & \textbf{4.5}  & -0.5  & \textbf{5.1}  & \textbf{4.7}  & 0.5   \\
  CO$_2$ Emission ($\downarrow$)    & 3.3  & 2.6  & -0.1  & 3.2  & 1.9  & 0.1   & 1.0  & -0.3 & -0.1  \\
  $\overline{\textnormal{TTC}}$ ($\downarrow$) & - & - & -1.2 & - & - & 1.0 & - & - & -1.2 \\
  TET ($\downarrow$)             & - & - & -1.2 & - & - & 1.2 & - & - & -1.2 \\ \hline
  \hline
  \multicolumn{10}{l}{The symbol ``-'' indicates that the corresponding values are close to zero.} \\
  \end{tabular}
\end{table*}

\begin{table}[h]
  \caption{Average number of lane changes per vehicle per evaluation episode under various demand conditions.\label{tab:train_lane_change_count}}
  \centering
  \renewcommand{\arraystretch}{1.3} 
  \begin{tabular}{p{0.55in}p{0.35in}p{0.6in}p{0.6in}p{0.55in}}
  \hline
  \hline
  \multirow{2}{*}{\textbf{Scenario}}        & \multirow{2}{*}{\textbf{Models}} & \multicolumn{3}{c}{\textbf{Demand}}  \\ \cline{3-5}
                                 &                         & Low   & High & Congested High    \\ \hline
  \multirow{2}{*}{\textit{Stable Flow}} & Baseline                   &  0.093$\pm$0.012   &  0.059$\pm$0.012  & 0.012$\pm$0.012   \\
                                 & Ours                & 0.086$\pm$0.011   & 0.036$\pm$0.008  & 0.009$\pm$0.009    \\ \hline
  \multirow{2}{*}{\textit{Lane Degrade}} & Baseline                   & 0.149$\pm$0.062   & 0.142$\pm$0.076  & 0.049$\pm$0.030   \\
                                 & Ours                & 0.106$\pm$0.043   & 0.057$\pm$0.021  & 0.043$\pm$0.027  \\ \hline
  \multirow{2}{*}{\textit{Vehicle Stop}} & Baseline                   & 0.143$\pm$0.058   & 0.135$\pm$0.061  & 0.061$\pm$0.036   \\
                                 & Ours              & 0.104$\pm$0.025   & 0.076$\pm$0.038  & 0.050$\pm$0.032   \\ \hline \hline
  \end{tabular}
  \end{table}

When energy cost is considered, a clear trade-off emerges between travel efficiency and energy cost in our model's performance. If the trade-off ratio between these two metrics is set to 1:1, our model performs better under high demand conditions compared to low demand conditions. For example, in the \textit{Lane Degrade} scenario under low demand conditions, our model achieves a {4.4}\% improvement in average speed at the cost of a {3.2}\% increase in CO$_2$ emissions. In contrast, under high demand conditions in the same scenario, these values are {4.5}\% and {1.9}\%, respectively. Considering the performance uplift margin, it is crucial to assess whether our model consistently achieves these gains. The environment is constructed to ensure fairness in comparison for our model and the baseline by maintaining the same vehicle route demand, safety-critical event settings, and random seeds in each evaluation round. 

\begin{figure*}[t]
\centering
\subfloat{\includegraphics[width=2in]{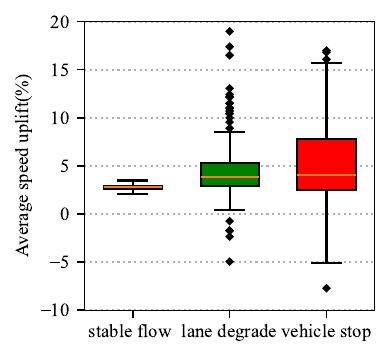}%
}
\subfloat{\includegraphics[width=2in]{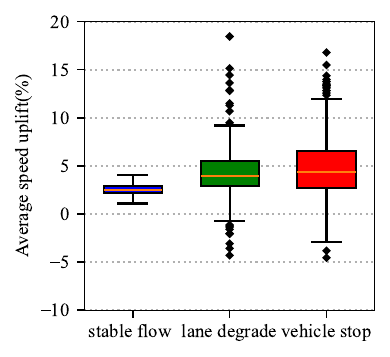}%
}
\subfloat{\includegraphics[width=2in]{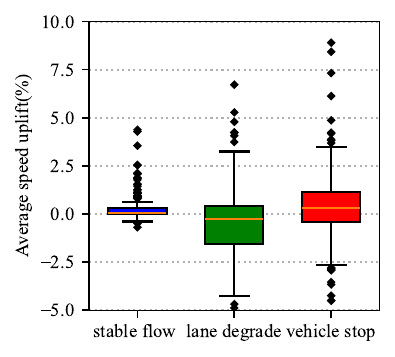}%
}
\caption{Boxplots of performance uplift across episodes under varied demand conditions. Each environment is evaluated over 200 episodes. From left to right, the plots represent low, high, and congested high demand conditions. The colored boxes represent the interquartile range (IQR) containing 50\% of the data near the median. Outliers are shown as black diamonds.}
\label{fig:eval_015_diff_model}
\end{figure*}

Fig. \ref{fig:eval_015_diff_model} shows the distribution of performance uplift across episodes in all environments. In low and high demand conditions, where our model outperforms the baseline, performance uplift is consistent across episodes. The \textit{Stable Flow} scenario exhibits less fluctuation than the \textit{Lane Degrade} and \textit{Vehicle Stop} scenarios, where fluctuations are primarily attributed to safety-critical events. This observation aligns with the higher fluctuation in training rewards under scenarios involving safety-critical events, as shown in Fig. \ref{fig:train_reward}. 

{Regarding safety metrics $\overline{\textnormal{TTC}}$ and TET, the values under low and high demand conditions are close to zero for both the baseline and our model under the TTC$^*=1.5$s threshold. Under congested high demand conditions, where TTC$^*=5.0$s is used, the changes remain small, within approximately $\pm 1.2\%$ across all scenarios. Although no safety-related term is included in the reward function, the model does not introduce evident safety degradation in the evaluated scenarios. This indicates that the proposed regulation framework can improve traffic efficiency while maintaining comparable safety performance in the microscopic simulation.}
\subsection{Policy Behavior Analysis}
Here, we analyze how our model optimizes traffic efficiency through lane-change regulation. The SL2015 lane-change model used in our experiments closely aligns with driver lane-change behaviors in real-world freeways, as described in \cite{erdmann2015sumo}. Specifically, the real-world regulation "Keep to the Right" is represented in SL2015 by the "Keep Right" lane-change intention, where vehicles are obligated to clear the overtaking lane. Additionally, lane changes based on the "Speed Gain" intention are accurately modeled in SL2015, capturing expected speed gains exceeding an empirical threshold. By default, decisions to speed up by changing to the right lane require more deliberation compared to leftward lane changes. To evaluate the policy behavior, we visualize the distribution of actions taken by our model in the \textit{Stable Flow} scenario under low and high demand conditions in Fig. \ref{fig:eval_action_rate}. We compare the action distribution at training checkpoints 100k and 200k for each evaluation environment and highlight two observations:

\begin{figure*}[t]
\centering
\subfloat{\includegraphics[width=3in]{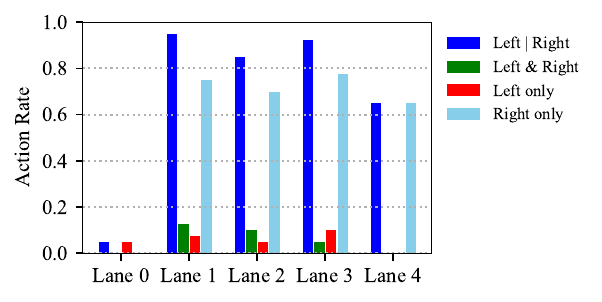}
}
\subfloat{\includegraphics[width=3in]{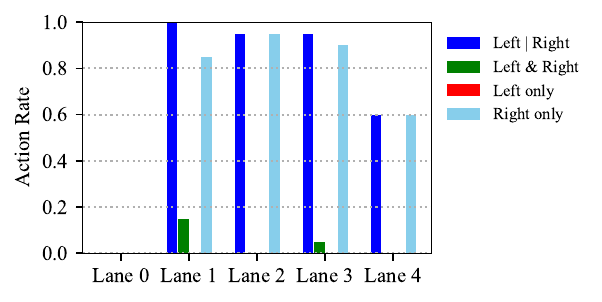}
} \\
\vspace{-1em}
\subfloat{\includegraphics[width=3in]{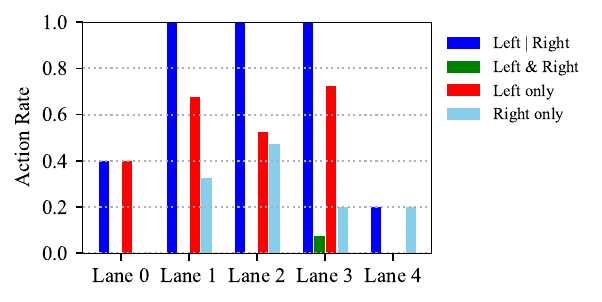}
}
\subfloat{\includegraphics[width=3in]{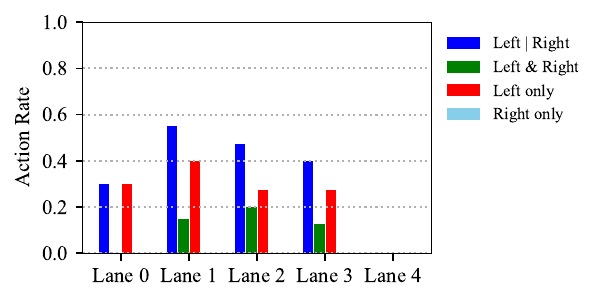}
} \\
\begin{minipage}[t]{3in}
\hspace{4em}
\centering \footnotesize Low Demand
\end{minipage}%
\hfill
\begin{minipage}[t]{3in}
\hspace{-10em}
\centering \footnotesize High Demand
\end{minipage} \\
\caption{Distribution of actions across different training checkpoints for the \textit{Stable Flow} scenario under low and high demand conditions. The training steps for each checkpoint, from top to bottom, are 100k and 200k steps. The lane index starts from 0, corresponding to the rightmost lane. Actions are collected every 1000 steps during evaluation for all lane grids along each lane.}
\label{fig:eval_action_rate}
\end{figure*}

\begin{enumerate}
\item The policy remains relatively permissive on the middle lanes. The rate of allowing at least one lane-change direction, denoted by '$\text{left} \mid \text{right}$' stays above 80\% under low demand and above 40\% under high demand. This indicates that the model does not simply prohibit all lane changes, but learns different regulation patterns under different demand conditions.

\item The learned action distribution shows clear demand-dependent behavior. Under high demand, the policy rarely selects the ``right only'' action. In addition, on the middle lanes, the rate of allowing at least one lane-change direction gradually decreases from the right-side middle lanes to the left-side middle lanes under high demand, while it remains close to 100\% under low demand. This pattern is consistent with the sparse traffic state under low demand and the stronger need for lane-change regulation under high demand.

\end{enumerate}

\begin{table}[t]
  \caption{Cross-evaluation results of the model trained in the \textit{Stable Flow} scenario applied to other scenarios.\label{tab:performance_uplift_cross_evaluate}}
  \centering
  \renewcommand{\arraystretch}{1.3} 
  \begin{tabular}{p{1in}p{0.4in}p{0.4in}p{0.4in}p{0.4in}}
  \hline
  \hline
  \multirow{2}{*}{\textbf{Metric}} & \multicolumn{2}{c}{\textbf{Lane Degrade}} & \multicolumn{2}{c}{\textbf{Vehicle Stop}} \\ \cmidrule(lr){2-3} \cmidrule(lr){4-5}
                                    & Low  & High  & Low  & High  \\ \hline
  Average Speed ($\uparrow$)     & 4.4/4.2$^*$  & 4.5/2.9     & 5.1/4.1  & 4.7/2.7     \\
  CO$_2$ Emission ($\downarrow$)    & 3.2/3.0      & 1.9/2.2     & 1.0/1.8  & -0.3/1.0    \\ \hline
  \hline
  \multicolumn{5}{l}{ *The value on the left is from Table \ref{tab:performance_uplift_all}, while the number on the right is } \\
  \multicolumn{5}{l}{from model cross validation.} \\
  \end{tabular}
\end{table}

Through this analysis, we further evaluate the performance of the policy trained in the \textit{Stable Flow} scenario when applied to other scenarios. As shown in Table \ref{tab:performance_uplift_cross_evaluate}, 
{cross-validation performance deteriorates under high demand conditions while remaining relatively close under low demand conditions. Under low demand, the average speed improvement changes from 4.4\% to 4.2\% in the \textit{Lane Degrade} scenario and from 5.1\% to 4.1\% in the \textit{Vehicle Stop} scenario. Under high demand, the deterioration is more evident. For example, the uplift in average speed decreases by 35.6\%, from 4.5\% to 2.9\%, in the \textit{Lane Degrade} scenario.}
This performance discrepancy can be attributed to the vehicle density differences between the two demand conditions. The sparse vehicle distribution under low demand conditions mitigates the impact of safety-critical events, whereas out-of-distribution congested traffic states induced by safety-critical events under high demand conditions are the primary cause of poor performance.

\subsection{Performance Comparison with PPO}

Previous work \cite{siboo2023empirical} investigates the performance of the off-policy Deterministic Policy Gradient (DDPG) and the on-policy PPO algorithm in an autonomous driving task with a continuous action space. The authors conclude that while PPO achieves competitive performance within a short training period, the off-policy DDPG algorithm performs better after convergence.
In our setting, although the action space is discrete, PPO remains a suitable baseline for performance comparison. We evaluate both PPO and Double DQN at the same training checkpoint where PPO demonstrates convergence, and also include the performance of the converged Double DQN model from Section~\ref{sec:result_performance}. The results are summarized in Table~\ref{tab:performance_uplift_vs_ppo}.

\begin{table*}[t]
  \caption{Comparison of performance metrics (average speed / CO$_2$ emissions) between Double DQN and PPO.\label{tab:performance_uplift_vs_ppo}}
  \centering
  \renewcommand{\arraystretch}{1.4} 
  \begin{tabular}{p{1.5in}p{0.3in}p{0.3in}p{0.3in}p{0.35in}p{0.35in}p{0.35in}}
  \hline
  \hline
  \multirow{2}{*}{\shortstack{\textbf{Algorithms} \\ \textbf{(state, checkpoints)}}} & \multicolumn{2}{c}{\textbf{Stable Flow}} & \multicolumn{2}{c}{\textbf{Lane Degrade}} &
  \multicolumn{2}{c}{\textbf{Vehicle Stop}} \\ \cmidrule(lr){2-3} \cmidrule(lr){4-5} \cmidrule(lr){6-7}
                                    & Low  & High   & Low  & High   & Low  & High   \\ \hline
  PPO (converged, $150k$)    & 0.0/0.0  & 3.0/2.5      & 4.3/3.2  & 0.3/-1.1     & 0.0/0.0  & 4.3/-0.4    \\
  DDQN  (not converged, $150k$)    & 2.1/2.7  & 1.2/1.7       & 2.5/2.3  & 3.6/1.9        & 4.8/1.4  & 1.3/-1.8    \\
  DDQN  (converged, $200k$)    & 2.8/3.3  & 2.6/2.6      & 4.4/3.2  & 4.5/1.9       & 5.1/1.0  & 4.7/-0.3    \\
  \hline
  \multicolumn{7}{l}{All values are percentage changes compared with the uncontrolled baseline in each environment.} \\
  \end{tabular}
  \end{table*}

{
PPO exhibits faster convergence and better sample efficiency at the evaluated $150k$ checkpoint. At this intermediate training stage, PPO performs competitively in several non-congested cases, such as the \textit{Stable Flow} scenario under high demand and the \textit{Lane Degrade} scenario under low demand. However, its performance is less stable across scenarios. For example, in the \textit{Stable Flow} and \textit{Vehicle Stop} scenarios under low demand, PPO converges to a dummy policy where all lane changes are allowed, resulting in no improvement over the uncontrolled baseline. In addition, PPO achieves only a 0.3\% speed improvement in the \textit{Lane Degrade} scenario under high demand.

Compared with PPO and the intermediate Double DQN checkpoint, the converged Double DQN checkpoint provides more consistent improvements across the evaluated scenarios. Given the relatively small discrete action space in our lane-change regulation problem, Double DQN proves to be an effective choice for training the proposed policy, offering stable performance with an acceptable training cost.
}

\subsection{Performance under Varying Penetration Rate of CVs} \label{sec:result_diff_rate}

\begin{figure}[t]
\centering
\subfloat{\includegraphics[width=3in]{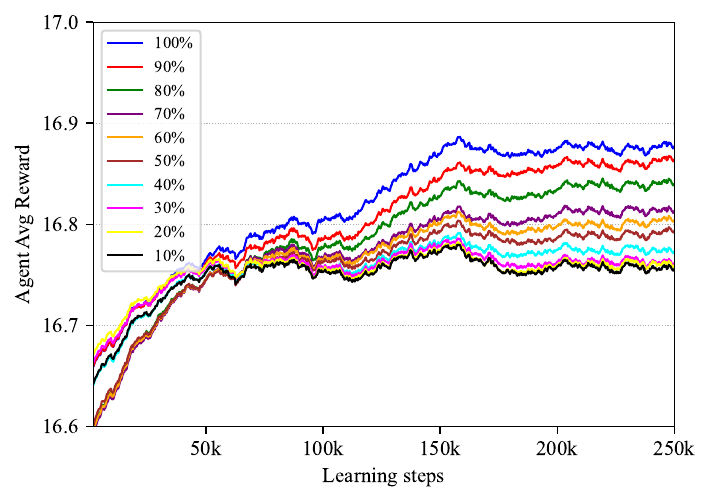}
}\\
\vspace{-0.5em}
\subfloat{\includegraphics[width=3in]{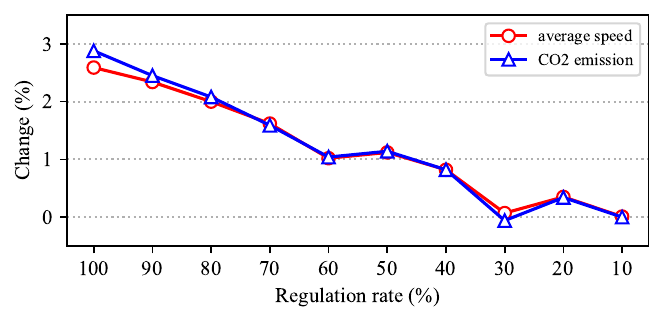}
}\\
\vspace{-0.5em}
\caption{Training reward and evaluation performance under varying regulation rates. Top: average training reward per episode. Bottom: evaluation results in terms of travel speed and energy cost.}
\label{fig:eval_control_rate}
\end{figure}

In previous experiments, all vehicles were {subject to regulation} (i.e., 100\% of vehicles were CVs). Here, we investigate how the performance of our model changes under different CV penetration rates. For each specified regulation rate, vehicles are randomly assigned as CVs in proportion to that rate, and the model is trained and evaluated accordingly. We select the \textit{Stable Flow} scenario under low demand conditions to examine the effect of regulation rate on model performance. Fig.~\ref{fig:eval_control_rate} presents the training rewards and evaluation metrics under varying CV penetration rates.

As the regulation rate decreases from 100\% to 10\%, the average agent reward per episode during training also declines. Evaluation results from converged checkpoints show that both average travel speed and CO$_2$ emissions decrease as the regulation rate drops. In Section~\ref{sec:result_performance}, we observed that the efficiency gain in speed closely balances the cost increase in emissions for the \textit{Stable Flow} scenario under low demand. This conclusion remains valid for partial regulation, as long as the CV penetration rate exceeds a certain threshold.

When the regulation rate falls below 30\%, the model's performance becomes comparable to the uncontrolled baseline. This suggests that a minimum level of regulation is necessary to activate the benefits of our framework. Once this requirement is met, our model continues to function effectively even when full lane change regulation is not accessible.

\section{Conclusion} \label{sec:conclusion}
This paper presents a novel lane change regulation model designed to enhance overall traffic efficiency. The approach introduces the concept of regulating individual connected vehicle lane change decisions through a macroscopic traffic control framework. A multi-agent reinforcement learning algorithm is proposed, wherein the action space determines whether to permit or prohibit lane change actions within each lane grid. Simulation environments are constructed across various scenarios and demand conditions to comprehensively evaluate the performance of the proposed model.

In scenarios {with 100\% CV penetration}, experimental results demonstrate that our model achieves comparable or superior performance relative to the baseline when balancing traffic efficiency and energy cost.
{Performance analysis reveals that the model learns demand-dependent and lane-dependent regulation patterns rather than simply prohibiting all lane changes.}
Additionally, experiments with varying regulation rates indicate that the model continues to positively influence traffic with the control rate above a certain threshold.

Three promising directions for future research are identified.
First, the current model assumes that all connected vehicles fully comply with regulation signals. However, individual objectives may conflict with the global goal, potentially leading to regulation violations. To address this, future work can incorporate a high-fidelity simulator that models human compliance behavior.
Since cross-environment experiments reveal performance degradation, exploring methods to train a unified, generalized policy adaptable to varied demand conditions and scenarios deserves investigation.
Finally, real-world deployment requires bridging the simulation-to-reality gap. Future work should generalize to a wider range of real-world demand conditions and scenarios. 
\bibliographystyle{IEEEtran}
\bibliography{IEEEabrv, references}

\end{document}